\documentclass[12pt]{article}

\usepackage[reqno]{amsmath}
\usepackage{mathrsfs}
\usepackage{amssymb}

\usepackage{rotating} 

\usepackage{array}
\usepackage{float}
\usepackage{color}
\usepackage{graphicx}

\usepackage{a4}
\usepackage{a4wide}
\usepackage{wasysym}

\def\gs{\mathrel{
   \rlap{\raise 0.511ex \hbox{$>$}}{\lower 0.511ex \hbox{$\sim$}}}}
\def\ls{\mathrel{
   \rlap{\raise 0.511ex \hbox{$<$}}{\lower 0.511ex \hbox{$\sim$}}}}

\newcommand{\be}{\begin{eqnarray}}
\newcommand{\ee}{\end{eqnarray}}
\newcommand{\beq}{\begin{equation}}
\newcommand{\eeq}{\end{equation}}
\newcommand{\bena}{\begin{eqnarray}}
\newcommand{\eena}{\end{eqnarray}}

\begin{document}

\setlength{\unitlength}{1mm}

\begin{titlepage}
\title{\vspace*{-2.0cm}
\bf\Large
keV sterile Neutrino Dark Matter and Neutrino Model Building
\\[5mm]\ }

\author{
Alexander Merle\thanks{email: \tt amerle@kth.se}
\\ \\
{\normalsize \it Department of Theoretical Physics, School of Engineering Sciences,}\\
{\normalsize \it KTH Royal Institute of Technology, AlbaNova University Center,}\\
{\normalsize \it Roslagstullsbacken 21, 106 91 Stockholm, Sweden}
}
\date{\today}
\maketitle
\thispagestyle{empty}

\begin{abstract}
\noindent
A sterile neutrino with a mass around the keV scale could be an interesting candidate for warm dark matter. Although there are several scenarios and production mechanisms known in which such a particle could yield the correct abundance, there are astonishingly few models around that can actually yield an explanation for the appearance of a keV-like scale. We here review three main classes of such mass models for keV sterile neutrino dark matter, based on split seesaw, on $L_e-L_\mu-L_\tau$ symmetry, and on the Froggatt-Nielsen mechanism, respectively.
\end{abstract}

\end{titlepage}

\section{\label{sec:Intro}Introduction}

One hot topic in contemporary astroparticle physics is the question about the nature of the dark matter (DM) in the Universe. One can imagine different scenarios~\cite{Steffen:2008qp}: DM could be highly relativistic today (\emph{hot dark matter}, HDM), it could be non-relativistic (\emph{cold dark matter}, CDM), or in between (\emph{warm dark matter}, WDM). Out of these, CDM is considered to be the standard scenario~\cite{Komatsu:2010fb}, while HDM is ruled out as the dominant DM component by structure formation~\cite{DelPopolo:2008mr}. However, WDM is still compatible with computer simulations on structure formation~\cite{Bode:2000gq}, and also recent analyses of the current data seem to point towards WDM~\cite{deVega:2009ku,Papastergis:2011xe}.

A well-motivated candidate particle for WDM is a sterile neutrino with a mass at the keV scale: First, sterile (right-handed) neutrinos are required in many models for a non-zero neutrino mass, and second, the mass of the sterile neutrinos is generated by a Majorana mass term which is \emph{not} bound to the electroweak scale. Hence, a sterile neutrino could very well have a mass of a few keV. This immediately suggests interesting connections to neutrino model building.

Before discussing some model building aspects, i.e., how to find an explanation for the appearance of a keV-scale mass, let us shortly comment on a few \emph{production mechanisms} of keV sterile neutrinos. It is clear that, in order to constitute the DM today, the WDM candidates have to be produced in the right amount in the early Universe. Indeed, there are several production mechanisms available as, e.g., non-thermal production as used in the $\nu$MSM~\cite{Asaka:2005an,Asaka:2006nq}, by having a primordial abundance (e.g.\ due to inflaton decay)~\cite{Anisimov:2008qs,Bezrukov:2009yw}, or by thermal overproduction and subsequent dilution by entropy production~\cite{Bezrukov:2009th}. On the other hand, it is required to find a dedicated explanation for the keV-scale itself, since it is different from other ``natural'' scales in particle physics. While we focus on such \emph{mass mechanisms} here, it is the ultimate goal to find models containing explanation for both, the keV scale as well as the WDM production mechanism.

\section{\label{sec:models}Mass models for keV sterile neutrino DM}

Note that, in this section, when we talk about \emph{mass mechanisms}, then this refers to possible models that indeed give (to some extent) an explanation for a keV-like mass, in the form of, e.g., a suppression mechanism or a suitable form of the mass spectrum. Relating certain quantities and thereby explaining patterns is the maximum one can do, but at the moment there is no method known to predict absolute scales. Contrary to mass mechanisms, there are also \emph{production mechanisms} (sometimes also called ``models''), a term we use to refer to settings that include keV sterile neutrinos without providing an explanation for the keV scale. Scenarios containing production mechanisms are extremely useful for concrete calculations, while models containing mass mechanisms try to explain the scales involved.

\subsection{\label{sec:Split}Split seesaw}

The first mass mechanism discussed here is based on the \emph{split seesaw mechanism}~\cite{Kusenko:2010ik}. This mechanism is very powerful in suppressing and separating scales, due to an exponential factor involved. The trick is to consider a 5-dimensional theory that is compactified on a $S^1/Z_2$ manifold, with a 5D coordinate $y\in [0,l]$ which separates the Standard Model (SM) brane at $y=0$ from the ultraviolet brane at $y=l$. The right-handed (RH) neutrinos can, as SM singlets, penetrate the bulk and hence their zero modes acquire a wave function the depends exponentially on $y$:
\begin{equation}
 \Psi_{R i}^{(0)}=\sqrt{\frac{2 m_i}{e^{2 m_i l}-1}} \frac{1}{\sqrt{M}} e^{m_i y} \Psi_R^{\rm 4D}(x)\ \Rightarrow\ M_{R i}= \kappa_i v_{B-L} \frac{2 m_i / M}{e^{2 m_i l}-1}.
 \label{eq:split_1}
\end{equation}
It is exactly this exponential profile that leads to an exponential suppression of the 4D-masses of the zero models! Furthermore, the 5D masses $\{ m_i \}$ can be relatively close together, e.g., a moderate splitting of $(m_1,m_2)\simeq (24,2.3) l^{-1}$ is enough to achieve an extreme splitting of the physical masses, $(M_1,M_2)\sim ({\rm keV}, 10^{12}{\rm GeV})$, which exemplifies the strength of this mechanism. Note that this mechanism can be refined using a flavour symmetry, such as $A_4$~\cite{Adulpravitchai:2011rq}: The symmetry can be used to generate the small splitting of the 5D masses in the first place, which is then used as ``input'' spectrum for the split seesaw.

One further bonus of this model is that the seesaw mechanism is guaranteed to work: Due to the Yukawa couplings receiving similar exponential suppression factors, the exponentials actually drop out of the seesaw formula. Hence, in the seesaw, the neutrinos only ``feel'' the natural (high) scale $v_{B-L}$ of the right-handed neutrino masses, and there arises no problem due to the appearance of a low keV scale in the seesaw denominator.

\subsection{\label{sec:Le}$L_e-L_\mu-L_\tau$ symmetry}

A model that achieves a mass splitting by an $L_e-L_\mu-L_\tau$ flavour symmetry was first suggested in general by Ref.~\cite{Mohapatra:2001} and then applied to the keV sterile neutrino case in~\cite{Shaposhnikov:2006nn}. Later on, this independently worked out in detail in~\cite{Lindner:2010wr}: As known from earlier studies~\cite{Lavoura:2000ci}, exact $L_e-L_\mu-L_\tau$ symmetry leads to light neutrino mass patterns of the form $(0,m,m)$, which means that one neutrino is massless while the other two are degenerate. It was spotted in Ref.~\cite{Lindner:2010wr} that this idea could easily be extended to the RH neutrino sector, which yields a similar pattern $(0,M,M)$ for the heavy neutrino masses.

\begin{figure}[t]
\centering
\begin{tabular}{lr}
\includegraphics[width=5.5cm]{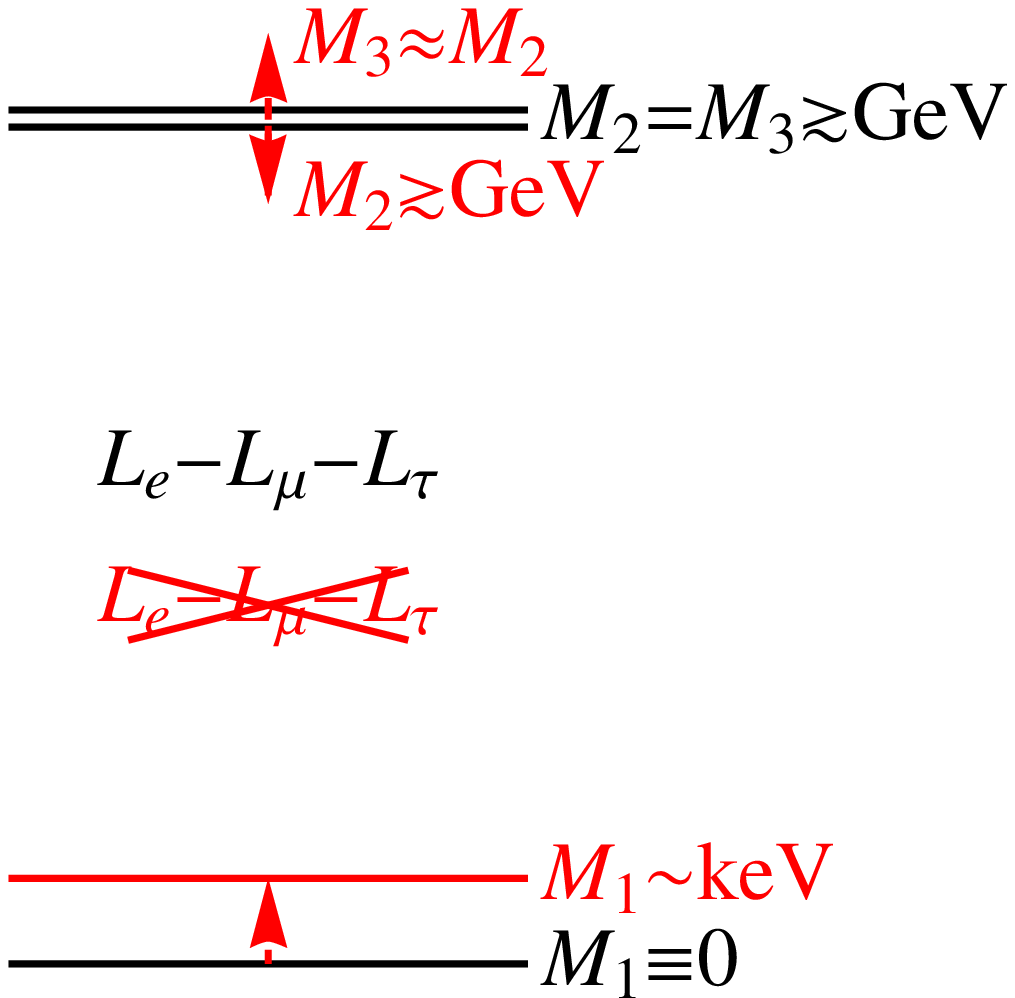}\hspace{0.5cm}  & \includegraphics[width=7.2cm]{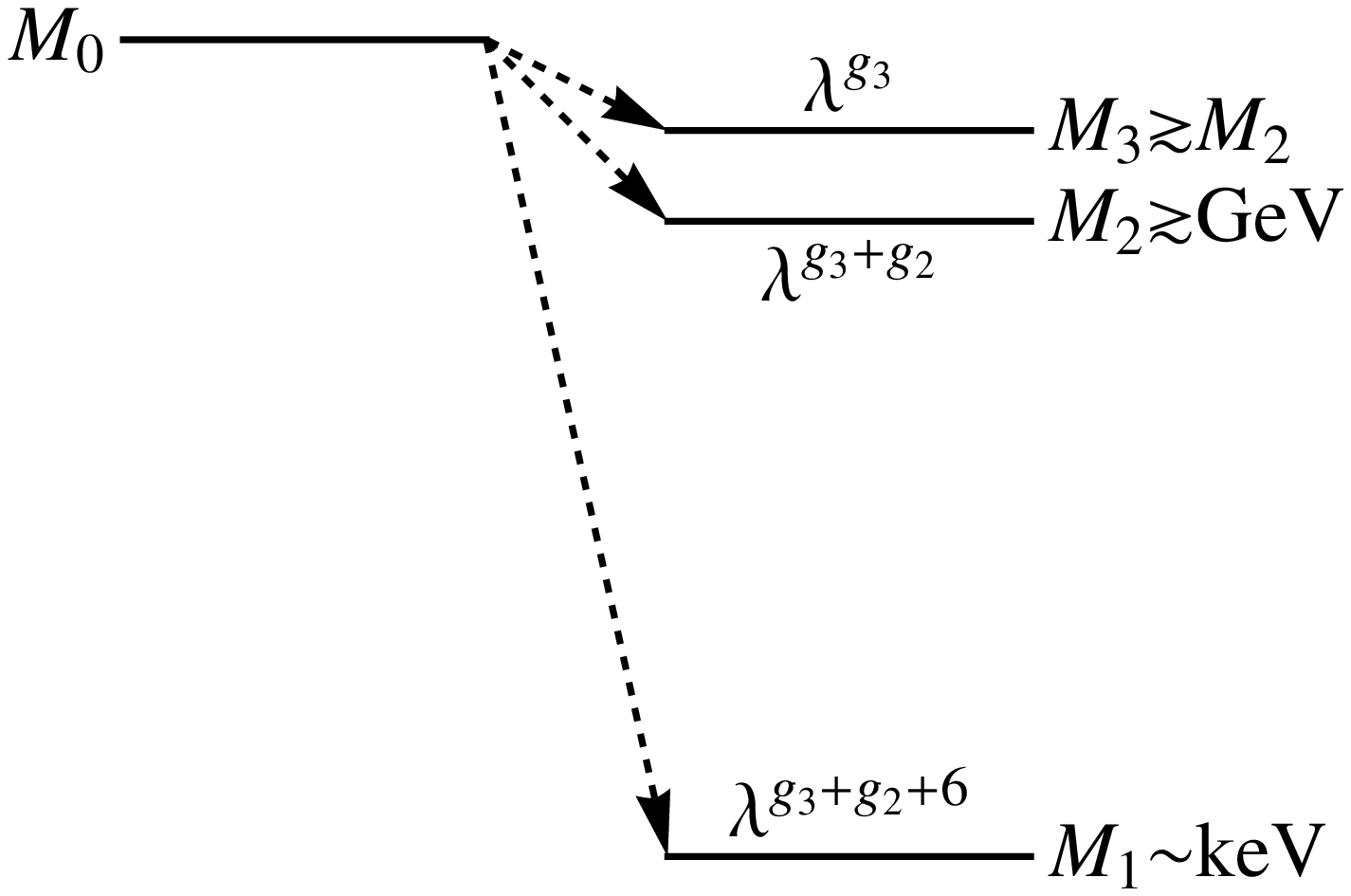}
\end{tabular}
\caption{\label{fig:schemes} The mass shifting schemes for the $L_e-L_\mu-L_\tau$ and the Froggatt-Nielsen models. (Figures taken from Ref.~\cite{Merle:2011yv}.)}
\end{figure}

A pattern like $(0,m,m)$ is not compatible with neutrino oscillation data, and hence the flavour symmetry $L_e-L_\mu-L_\tau$ must be broken. This breaking is often parametrized by so-called \emph{soft breaking terms}. These terms do not respect the initial flavour symmetry and they can be chosen more or less arbitrarily. The point is that any choice will have two generic effects, namely making the massless neutrino massive, by bringing in a new scale, while simultaneously lifting the degeneracy of the non-zero masses,
\begin{equation}
 (0,M,M) \to \left( \mathcal{O}(S), M-\mathcal{O}(S), M+\mathcal{O}(S) \right).
 \label{eq:Le_1}
\end{equation}
This is exemplified in the left panel of Fig.~\ref{fig:schemes}. Since the breaking scale must be small compared to the symmetry preserving scale, $S\ll M$, as otherwise we would not speak of a symmetry in the first place, this mechanism yields a perfect motivation for $S\sim {\rm keV}$, while $M\gtrsim \mathcal{O}({\rm GeV})$ or heavier.

\subsection{\label{sec:FN}The Froggatt-Nielsen mechanism}

The third model containing an explanation for the keV scale~\cite{Merle:2011yv} was based on the Froggatt-Nielsen (FN) mechanism~\cite{Froggatt:1978nt}. By introducing an unknown high energy sector, this mechanism leads to generation-dependent suppression factors in the mass matrices, see right panel of Fig.~\ref{fig:schemes}. It leads to an extremely strong (exponential) suppression of certain masses, and hence to the appearance of strong hierarchies. Thus, FN has become very popular for the explanation of quark mass patterns, but it is equally suited for keV sterile neutrinos, since a strong mass splitting is required by many production mechanisms.

Similar to what happens in the model presented in Sec.~\ref{sec:Split}, the FN mechanism guarantees seesaw to work. This is because any global $U(1)$ charges of the RH neutrinos, and in particular the one of $U(1)_{\rm FN}$, are forced to cancel in the seesaw mass matrix, thereby avoiding any small scales in the denominator. In addition it has been shown in Ref.~\cite{Merle:2011yv} that the FN mechanism, although usually involving a certain degree of arbitrariness, is actually quite constrained when applied to keV sterile neutrino scenarios. This allows for predictivity, as well as compatibility with only some frameworks. For example, an embedding in an $SO(10)$ Grand Unified Theory is strongly disfavoured as compared to $SU(5)$, and FN scenarios are incompatible with Left-Right symmetry, which was the main example used in the production mechanism of Ref.~\cite{Bezrukov:2009th}. In turn, models based on the FN mechanism are falsifiable not only by more information on low energy neutrino data, but also by verifying one of the settings and/or production mechanisms which are incompatible with FN.

\section{\label{sec:conc}Conclusions}

A sterile neutrino with a mass at the keV scale is a very interesting candidate for WDM, and its existence is well motivated from particle physics, too. Although many scenarios and production mechanisms have been discussed in the literature, there is a lack of models than can explain the appearance of the keV scale in the first place. This gap is currently being filled, and three such examples have been presented here. We can hopefully look forward to many more interesting ideas for new mass and production mechanisms of keV sterile neutrino dark matter.

\section*{Acknowledgements}

AM thanks the organizers of the TAUP conference 2011 for doing an excellent job, and he also thanks all the participants he could share interesting discussions with. Furthermore, AM is particularly grateful to his collaborators M.~Lindner and V.~Niro. The work of AM is supported by the G\"oran Gustafsson foundation.

\end{document}